\documentclass[doublecolumn]{aa} 

\usepackage{afterpage}
\usepackage{graphicx}
\usepackage{txfonts}
\usepackage{wasysym}
\usepackage{siunitx}
\usepackage{placeins}
\usepackage{txfonts}
\usepackage{booktabs}
\usepackage{fontawesome}
\usepackage[colorlinks=true,linkcolor=blue,citecolor=blue]{hyperref}%
\usepackage{color}
\usepackage{listings}
\usepackage[hyphenbreaks]{breakurl}

\begin{document}

   \title{ExoNAMD: Leveraging the spin-orbit angle to constrain the dynamics of multiplanetary systems}

   \subtitle{}

\author{
    A. Bocchieri\inst{1, 2}\and 
    J. Zak\inst{3} \and
    D. Turrini\inst{4}
}

\institute{
Dipartimento di Fisica, La Sapienza Università di Roma, Piazzale Aldo Moro 5, Roma, 00185, Italy\\\email{andrea.bocchieri@uniroma1.it}\and 
INAF - Osservatorio Astrofisico di Arcetri, Largo Enrico Fermi 5, 50125, Firenze, Italy \and 
Astronomical Institute of the Czech Academy of Sciences, Fri\v{c}ova 298, 25165 Ond\v{r}ejov, Czech Republic \and  
INAF - Osservatorio Astrofisico di Torino, Via Osservatorio 20, I-10025, Pino Torinese, Italy
}

   \date{Received April XX, 2024; accepted April XX, 2024}

  \abstract{Multiplanetary systems are excellent laboratories for studying the formation and evolution of exoplanets inside the same stellar environment. The number of known multiplanetary systems is expected to skyrocket with the advent of \textit{PLATO} and the Roman space telescope. The spin-orbit angle is a key context information for the systems' dynamical history, and in recent years a growing number of planets had their spin-orbit angle measured, revealing a large diversity in orbital configurations, from well-aligned to polar, and even retrograde, orbits. 
  Still, observers lack a robust tool to compare the dynamical state of different systems and to select the most suitable ones for future avenues of exploration, such as investigating the evolutionary pathways and their links to the atmospheric composition. 
  Here, we present \texttt{ExoNAMD}, an open source code aimed at evaluating the dynamical state of multiplanetary systems via the Normalized Angular Momentum Deficit (NAMD) metric. The NAMD measures the deficit in angular momentum with respect to circular, co-planar orbits. It is normalized to compare systems with different architectures and provides a lower limit on the past dynamical excitation of the system. We find that using the spin-orbit angle parameter in the NAMD calculation (A-NAMD) improves the dynamical state's description, compared to using only the relative inclinations (R-NAMD). Comparison of A-NAMD and R-NAMD also yields powerful insights into the interplay between eccentricity and spin-orbit angle. 
  \texttt{ExoNAMD} is a timely tool for easy and fast comparison of the myriad of exoplanetary systems to be discovered by \textit{PLATO} and Roman, and to optimize the target selection and scientific output for future atmospheric characterization using ELTs, JWST, and Ariel.}

   \keywords{Planets and satellites: dynamical evolution and stability -- Celestial mechanics -- Planet-star interactions -- Planets and satellites: atmospheres}
   \maketitle
%

\section{Introduction}
\subsection{Multiplanetary systems}

The discovery of about 6000 exoplanets in nearly 30 years has revolutionized our understanding of planets and planetary systems, previously based on the sole example of the Solar System. The discoveries reveal a vast diversity of physical and orbital properties of planets around different stellar types, from hot Jupiters to super-Earths, from around B-type stars to brown dwarfs and even pulsars. Recently, the emphasis in the field has shifted from discovery to detailed characterization. The parameter space to characterize, however, is highly multidimensional, posing a significant challenge to achieve a unified theory for planetary formation and evolution histories as for stars. 

Constraining the origins of individual exoplanets is a challenging task as different formation and migration histories can result in the same observed outcome \citep{daw18}. For example, the migration pathways of hot and warm Jupiters are still debated, with the contribution of gas disk migration \citep{ida04,bar14} vs. high-eccentricity migration via the Kozai-Lidov mechanism \citep{fab14} still unconstrained by observations. In this context, multiplanetary systems are crucial case studies, as their planets have formed and migrated under the same stellar environment. 

The observed population of multiplanetary systems ($\sim$\,1\,000 known systems) shows a great diversity \citep{mish23} covering a range of eccentricities, masses, orbital periods, and spin-orbit angles. No individual parameter nor single combination of parameters (metric) is sufficient to capture this diversity; specialized metrics need to be developed to homogeneously compare the properties of different systems, constrain their formation pathways, and select the most suitable for follow-up observations. The ultimate goal is to have a comprehensive theory of planets and how they formed and evolved, which allows us to assess the uniqueness of our own Solar System and conditions for life elsewhere in the Galaxy. 

There are two main methods for investigating the histories of planetary systems based on their present properties. The first is the spectroscopic characterization of exoplanetary atmospheres \citep{tin13,madhu19,tur22} that carry information about their formation and subsequent evolution.
The interpretation of spectroscopic data to constrain the planetary histories must account for how the atmospheric composition of planets can change over time.
Atmospheres connect the planet to its host star, as their upper layers are shaped by the stellar environment, which alters the composition via photochemistry \citep{for20,tsai23}. 
In multiplanetary systems, planetesimal bombardments and planetary collisions can alter or cancel the original signatures of planet formation~\citep{tur15, sel22, sain24, poly25}.
Furthermore, multiplanetary systems are excellent laboratories to test how these processes alter the original composition as a function of the orbital distance \citep{barat24}. The advent of JWST has transformed our view of exoplanetary atmospheres, shifting the field from detection of individual chemical species and partial classification to detailed characterization. Outstanding results include the robust detection of major trace gases such as CO$_2$, H$_2$O, and CH$_4$, as well as photochemically produced SO$_2$~\citep{rust23,tsai23,po24}. JWST successes also highlight three critical challenges and opportunities that the exoplanetary community will face in the next decades: (i) selecting optimal targets for detailed spectroscopic characterization with existing and future instrumentation; (ii) compiling unbiased target lists for large spectroscopic survey missions such as Ariel; (iii) correctly interpreting the atmospheric data~\citep[e.g., Three Key Criteria;][]{seag25}.

The second method to investigate planetary systems' histories is to characterize the architectures of the systems through the physical and orbital parameters of the planets and their host star \citep[e.g.,][]{tur22,leleu24}. Among these parameters, the spin-orbit angle\footnote{Each planet has its own spin-orbit angle, while the stellar obliquity parameter is given by the angular momentum of all the planets in the system. Previously, these parameters have been used interchangeably, but for multiplanetary systems they can be different.} (the physical equivalent of the absolute orbital inclination in celestial mechanics) has emerged in recent years as a powerful probe of the dynamical history. It is defined as the angle between the stellar spin axis and the normal to the orbital plane of the planet \citep{alb22}. The spin-orbit angle is a physical property intrinsic to each planet, unlike the inclination with respect to the celestial sphere, which depends on the line of sight. Recent measurements of spin-orbit angles have revealed an unexpectedly wide variety of orbits, from well-aligned to polar and even retrograde. 
Also in this case, multiplanetary systems offer a unique opportunity to study the phenomena behind these diverse architectures, dictated by the delicate interplay of multiple processes such as the Kozai-Lidov mechanism, secular resonances, mean motion resonances, planet-planet scattering, and eccentricity cascade.

These two methods are complementary, each revealing different aspects of the nature of planets. Only by leveraging their combined usage can we overcome their individual limitations in constraining the planetary systems' histories.

This work focuses on the second method, and aims to construct robust metrics and diagnostic tools to characterize the dynamics of planetary systems and allow cross-systems comparisons. Previous works \citep{las97,las00,las17} introduced the Angular Momentum Deficit (AMD) metric -- which measures the deviation of planetary orbits from the circular and co-planar case -- to study the dynamical stability of the Solar System and later exoplanetary systems. The focus of the AMD is on constraining the future stability of the systems. The AMD offers a computationally efficient method, requiring no numerical integration and relying solely on derived orbital parameters. However, AMD does not provide a timescale for potential instability and neglects important dynamical phenomena such as mean-motion resonances, which can significantly influence system behavior. 

In parallel to AMD, several other metrics have been developed to predict the future stability of planetary systems to address the limitations of the AMD metric. These include MEGNO \citep[Mean Exponential Growth of Nearby Orbits,][]{cin00} and SPOCK \citep[Stability of Planetary Orbital Configurations Klassifier,][]{tam20}, each with distinct advantages and drawbacks. 

MEGNO evaluates stability through numerical integration and chaos indicators, capturing intricate dynamical interactions and providing direct physical insights. However, MEGNO is computationally intensive and typically integrates systems over a limited number of orbits, potentially overlooking long-term dynamical trends. In contrast, SPOCK leverages machine learning trained on extensive numerical simulations to quickly predict whether planetary systems are dynamically stable or unstable. Although SPOCK offers rapid stability assessments without the computational expense of numerical integration, it relies heavily on pre-trained datasets and lacks the direct interpretability provided by MEGNO or AMD. Also, SPOCK is designed only for systems with three and more planets, limiting its applicability. Recent studies \citep[e.g.,][]{gaj23} have utilized these stability metrics to assess the future stability of planetary systems and compare their predictive capabilities.

Building upon AMD, and focusing instead on the past dynamical histories, \citet{cham01} and \citet{tur20} (hereafter, CH1 and T20) developed the Normalized Angular Momentum Deficit (NAMD) metric, which allows for homogeneous quantitative comparisons of dynamical excitation across different multiplanetary systems, regardless of their specific orbital architectures. The NAMD normalization eliminates the dependency on the planetary masses and semi-major axes, overcoming the limitation of the AMD which could not compare systems with compact \citep[e.g., TRAPPIST-1,][]{gill17} and extended architectures \citep[e.g., HR 8799,][]{maro10}. The interpretation of the NAMD is straightforward: higher NAMD values indicate greater dynamical excitation and higher probabilities of past dynamical instabilities, making NAMD an effective ``thermometer'' for evaluating the past dynamical history of planetary systems \citep[][hereafter, T22]{tur22}.

While qualitative information on multi-planet systems can be derived from basic considerations on their orbital properties -- e.g., whether the planetary eccentricities cause the orbits to intersect or whether the planets have very different spin-orbit angles -- this approach does not allow for homogeneous and quantitative comparisons between systems with different architectures. 
As an example, this approach does not allow to easily assess whether a planetary system containing two Saturn-mass planets on eccentric orbits is more or less excited than a system containing a Jupiter-mass planet on a more mildly eccentric one.  
By tying the dynamical excitation of planetary systems to their angular momentum -- hence weighting eccentricity and inclination by the mass and semimajor axis of the planets -- the NAMD naturally allows for this kind of homogeneous quantitative comparison \citep{tur20,tur22} and, in combination with metrics such as AMD, MEGNO, and SPOCK, allows for the complete description of the past, present and future dynamical states of planetary systems.

\subsection{Need for a revised NAMD formulation}

The formulation of the NAMD by T20 uses the relative inclination with respect to the most massive planet in the system as a proxy of the spin-orbit angle.
However, the NAMD in this formulation does not consider orbital misalignment with the stellar spin axis, which is captured in the spin-orbit angle parameter, $\psi$. 
For instance, without including this key parameter, the NAMD cannot capture the difference in the dynamical state of two systems with transiting planets: one characterized by planets on aligned orbits (orbiting in the same direction as the stellar rotation), and one with planets on polar or even retrograde orbits. The latter is characterized by a higher degree of dynamical excitation.

The original NAMD formulation by T20 was well justified with only a handful of spin-orbit angle measurements available at the time. However, with the advancement of high-resolution spectrographs \citep{schw16,pepe21,sei22}, Rossiter-McLaughlin (R-M) effect measurements have increased dramatically in number and quality since T20 \citep{knud24,espi24,zak25}, thus increasing the number of planets for which we have the spin-orbit angle measured to the point where an updated NAMD formulation is warranted.

Currently, the projected spin-orbit angle is known for more than 300 planets and out of those around 10\% are in multiplanetary systems. This parameter is the spin-orbit angle projected onto our line of sight. 
On the other hand, the de-projected spin-orbit angle (hereafter, spin-orbit angle), is known for about 150 planets and around 15\% of those are in multiplanetary systems. 
With the upcoming launch of the \textit{PLATO} space mission \citep{rauer24}, the stellar inclination of a large number of stars will be measured thanks to its long-term observing strategy. 
This, combined with dedicated ground-based follow-ups\footnote{\url{https://stel.asu.cas.cz/plato/spectrograph.html}} providing the projected spin-orbit angle \citep[e.g.,][]{zak25ps}, will lead to even faster growth in spin-orbit angle measurements.

Therefore, here we update the relative inclination-based NAMD formulation by T20 (hereafter, R-NAMD) by replacing the relative inclination w.r.t. the most massive planet in the system with the spin-orbit angle parameter, which gives the spin-orbit angle-based NAMD (hereafter, A-NAMD). The number of spin-orbit angle measurements is still limited as they are observationally expensive and hence the R-NAMD remains a viable tool to characterize the dynamics of the exoplanetary population. Furthermore, as we present in this paper, the R-NAMD and A-NAMD metrics can be used jointly to increase their diagnostic power and derive more detailed constraints on the orbital dynamics.

To directly support the community, we have implemented both the R-NAMD and A-NAMD calculations in an open-source Python tool, ExoNAMD\footnote{\url{https://github.com/abocchieri/ExoNAMD}}. 
The dynamical context provided by the NAMD enables (1) cross-system dynamical state comparisons; (2) unbiased target selection for future observations; (3) comprehensive dynamical descriptions alongside stability metrics (AMD, MEGNO, SPOCK) in the forthcoming era of \textit{PLATO} and \textit{Ariel} and (4) to interpret measurements of planetary atmospheres, as migration history shapes their composition and thermal structure.

\section{Methods}
\subsection{NAMD}

The general formulation of the NAMD defined in CH1 and T20 is

\begin{equation}
\label{eq:namd-general}
\text{NAMD} = \frac{\sum_k m_k \sqrt{a_k} \left(1 - \sqrt{1 - e_k^2} \cos i_k \right)}{\sum_k m_k \sqrt{a_k}};
\end{equation}

where $m_k$ is the mass of the $k$-th planet, $a_k$ its semi-major axis, $e_k$ its eccentricity, and $i_k$ its inclination. As discussed in T20, the NAMD measures the dynamical excitation of planetary systems, which is tied to the occurrence and intensity of dynamical instabilities, hence the ``violence'' of the dynamical evolution (T20, T22). The in-plane excitation appears in the NAMD through the orbital eccentricities, while the out-of-plane excitation through the inclinations. 

In the case of exoplanetary systems, however, the inclination of the planets is referred to the local plane of the celestial sphere and not to a physically meaningful plane. In the original study by CH1, focused on the formation of the inner Solar System, the inclinations $i_k$ were implicitly assumed as referring to the midplane of the Solar Nebula. To circumvent this issue when the spin-orbit angles are unknown, \citet{zin17} and \citet{tur20} introduced the approach of adopting as the reference plane the orbital plane of the most massive planet in the system, based on the reasoning that this planet was plausibly the least excited and, therefore, the one closer to the original disk midplane.

This approach led to the formulation of the relative NAMD (R-NAMD), defined in T20 as:
\begin{equation}
\label{eq:namdr}
\text{R-NAMD} = \frac{\sum_k m_k \sqrt{a_k} \left(1 - \sqrt{1 - e_k^2} \cos i_{r,k} \right)}{\sum_k m_k \sqrt{a_k}}
\end{equation}
where $i_{r,k}$ is the inclination of the $k$-th planet with respect to the most massive planet in the system. The out-of-plane excitation in the R-NAMD is a lower limit to the real one, given that $i_{r,k}$ underestimates the true spin-orbit angles and the impact of specific viewing geometries (e.g., if the line of nodes is along the line of sight). Thus, the R-NAMD provides a lower limit to the dynamical excitation of the system.

In this work, we expand the application of the NAMD methodology to exoplanetary systems by introducing the absolute NAMD (A-NAMD) by replacing \(i_k\) and \(i_{r,k}\) with the spin-orbit angle, $\psi_k$, which gives:

\begin{equation}
\label{eq:namda}
\text{A-NAMD} = \frac{1}{2} \frac{\sum_k m_k \sqrt{a_k} \left(1 - \sqrt{1 - e_k^2} \cos \psi_k \right)}{\sum_k m_k \sqrt{a_k}}.
\end{equation}

The expression is divided by 2 to ensure that A-NAMD values are normalized to 1. The individual parameters needed for the R-NAMD and A-NAMD computation are briefly explained in the Appendix \ref{app:params}, together with the most widely-used techniques to obtain them. The A-NAMD provides a representative estimate of both in-plane and out-of-plane dynamical excitation. When the spin-orbit angle is known for only one planet in the system, this information can be used in principle to extrapolate the spin-orbit angles $\psi_{r,k}$ relative to the plane perpendicular to the stellar spin axis from the relative inclinations of the planets. As in the case of R-NAMD, however, this approach does not account for the impact of specific viewing geometries of the orbits.

NAMD is a temporally evolving quantity, as tidal forces can realign and circularize the orbits, lowering the dynamical excitation over time. This is true for both the R-NAMD and A-NAMD formulations. In Sec.~\ref{sec:discussion}, we discuss the relevant timescales for orbital circularization and realignment in the context of the R-NAMD and A-NAMD metrics.

\subsection{ExoNAMD: a community tool}

To date, there is no open-source tool available for the community to compute the NAMD, in either form. Hence, to fill this gap and streamline the preparation of follow-ups of multiplanetary systems, we present a new community tool called \texttt{ExoNAMD}, which we use for the analyses in this study. 

\texttt{ExoNAMD} is a user-friendly Python 3.8+ code that computes both the R-NAMD and A-NAMD and associated uncertainties, enabling quick and efficient comparisons of the remaining dynamical violence in the architecture of multiplanetary systems. 
The code is extensively documented on \textit{Read the Docs}\footnote{\url{https://exonamd.readthedocs.io/en/latest/}}, including examples and guides.
\texttt{ExoNAMD} is available on PyPI\footnote{\url{https://pypi.org/project/exonamd/}} for installation via 
\begin{lstlisting}[language=bash]
$ pip install exonamd
\end{lstlisting}

The steps to compute the NAMD are:
\begin{enumerate}
    \item Retrieve the planet's parameters, either from the NASA Exoplanet Archive\footnote{\url{https://exoplanetarchive.ipac.caltech.edu/}} or parsing a user input .csv file;
    \item Input any missing parameters and uncertainties (see the code documentation and/or Appendix \ref{app:missing});
    \item Assess the variability of the parameters via Monte-Carlo methods using their uncertainties;
    \item Compute the NAMD using the spin-orbit angle (A-NAMD) or the relative inclination w.r.t. the most massive planet (R-NAMD). 
\end{enumerate}

The R-NAMD offers insights on the systems' dynamics when the spin-orbit angle is not available. 
The A-NAMD allows for a more complete characterization of the system's dynamical state, as it takes into account its true architecture. See the following section for examples (such as K2-290 and Kepler-462) where only the inclusion of the $\psi$ parameter enables the correct interpretation of the system's architecture and possible planetary evolutionary pathways.

\subsection{Building the sample}
Here, we describe the steps used to obtain the sample of planetary systems we analyze in the following section. All these steps are natively implemented in \texttt{ExoNAMD}.

We query the NASA Exoplanet Archive for all multiplanetary systems and retrieve 1\,005 unique multiplanetary systems (date: 7th August 2025\footnote{During the first execution \texttt{ExoNAMD} queries the Archive on a given date; subsequent executions using the ``-u'' flag will update the database with the new entries. See the documentation for more details.}). We create a database of aliases for the removal of duplicates. We solve for missing values of stellar mass, orbital period or semi-major axis using the third Kepler law if any two parameters of this set of three are available. We also solve for the stellar radius or the semi-major axis if the database reports only their ratio, and similarly for the stellar radius and the planetary radius.
For each planet, we take the median value from all queried values and their uncertainties obtained from the Archive to avoid biases and outliers in the subsequent analysis, discarding any NaNs. Then, we clean our sample by discarding duplicate systems using our database of aliases, obtained using a custom catalogue implementation available in \texttt{ExoNAMD} which is inspired by the \texttt{gen\_tso} code \citep{cub24}. 

In the event that some of the parameters are not available, \texttt{ExoNAMD} offers the possibility to interpolate their values (see Appendix \ref{app:missing} for details). 

Following the methodology in T20, we compute both the R-NAMD and A-NAMD together with their uncertainties using a Monte Carlo approach. 
We use the \texttt{truncnorm} function from \texttt{scipy.stats} and we draw 100,000 random samples from a truncated normal distribution within the physical bounds of each parameter (e.g., non-negative masses). The distribution is centered at the obtained value from the previous steps and its half-width is obtained from the arithmetic mean of the upper and lower error bars.
We perform this step on all parameters needed for the R-NAMD and A-NAMD calculations in Eqs.~\ref{eq:namdr} and \ref{eq:namda}.
We take the median of the final distribution as the expected NAMD value, with associated uncertainties from the 16-84 percentile. 
Finally, we compute the relative uncertainty on the NAMD following T20: we take the half inter-quartile range and normalize it by the median. This allows us to keep track of the cases with too high relative uncertainty (higher than unity) when we interpret the results.

We also include the Solar System in the analysis as a reference point using the values reported on the NASA Planetary Fact Sheet\footnote{\url{https://nssdc.gsfc.nasa.gov/planetary/factsheet/}} following T20 and T22.

\section{Results}
\label{sec:results}

After performing the above steps, we are left with 792 multiplanetary systems, including those with interpolated values. Following T20, we decided to restrict our sample to those systems for which we have all the parameters for the R-NAMD calculation (without interpolations) to avoid possible biases.
We refer to this as the ``core'' sample and it consists of 67 systems.
The multiplicity distribution in the core sample is:

\begin{enumerate}
    \item 45 systems with $M$=2;
    \item 8 systems with $M$=3;
    \item 7 systems with $M$=4;
    \item 2 systems with $M$=5;
    \item 3 systems with $M$=6;
    \item 1 systems with $M$=7 (TRAPPIST-1);
    \item 1 system with $M$=8 (the Solar System).
\end{enumerate}

\begin{figure}[htbp]
    \centering
    \includegraphics[width=1\linewidth]{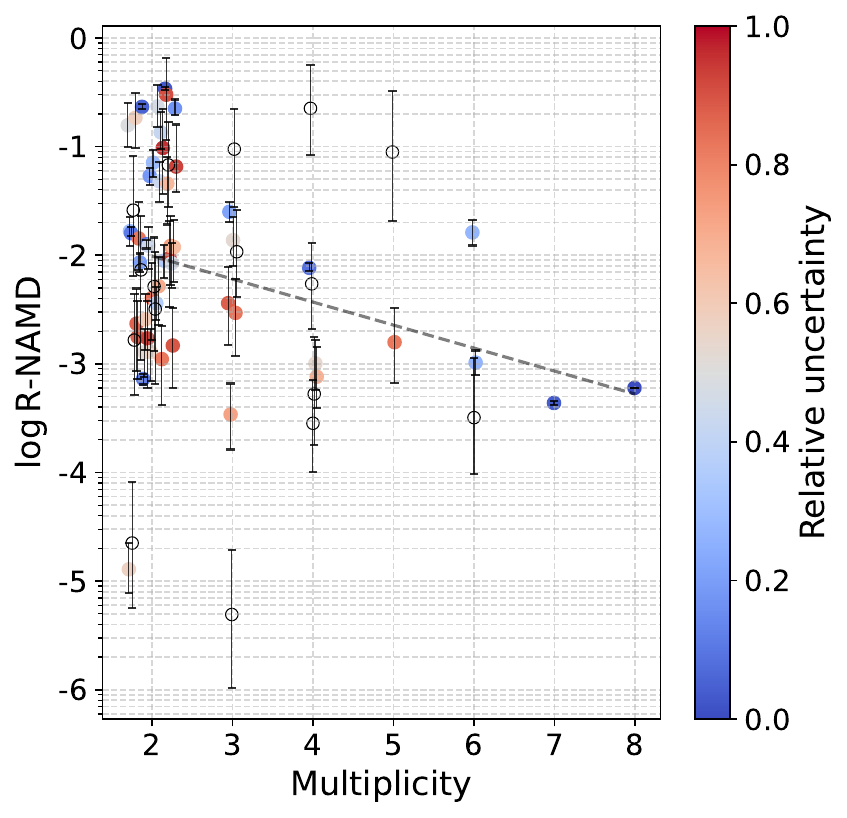}
    \caption{R-NAMD vs.~multiplicity plot for the core sample described in the text. Same figure as Fig.\,3 in T20 but with an expanded sample. The horizontal offsets around the multiplicity values are put arbitrarily for better visualization. The color scale, capped to 1, represents the relative uncertainty on the R-NAMD values, obtained from the arithmetic mean of the lower and upper errors from the Monte Carlo. Systems with relative uncertainty above unity are plotted with a white hollow circle. The linear fit is shown only as a visual aid as the slope is highly dependent on the (few) high multiplicity systems.}
    \label{fig:namdr-core-sample}
\end{figure}

\begin{figure}[!h]
    \centering
    \includegraphics[width=1\linewidth]{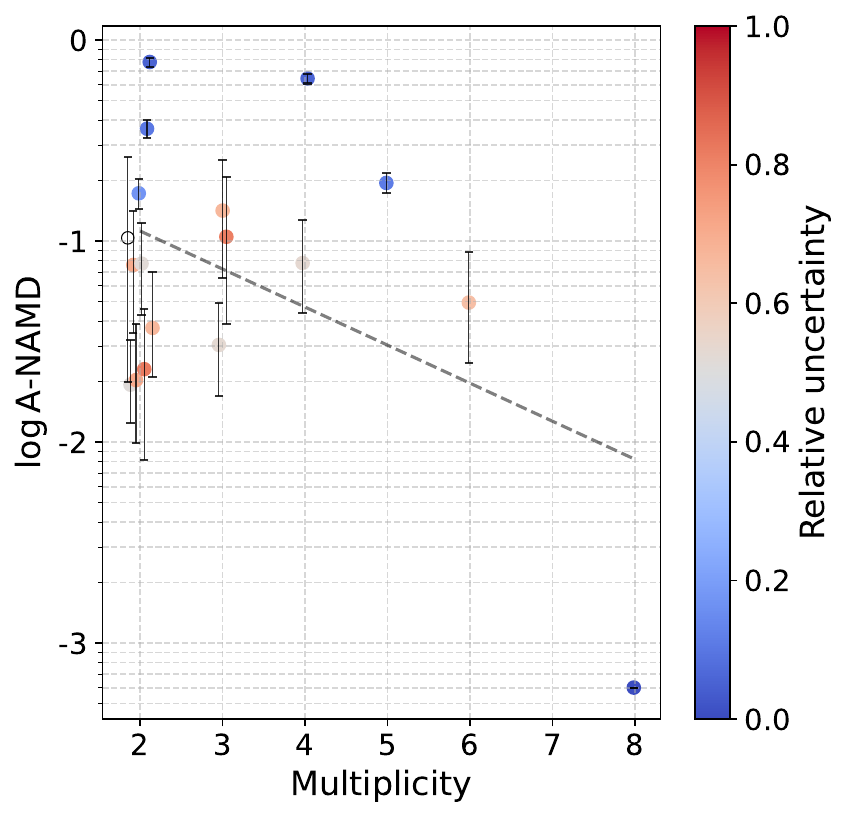}
    \caption{A-NAMD vs.~multiplicity plot. The shown systems are selected from the core sample on the basis of the availability of spin-orbit angle measurements, as described in the text. More systems are needed to confirm and quantify the sketched anti-correlation between A-NAMD and multiplicity. The linear fit is shown only as a visual aid.}
    \label{fig:namda-core-sample}
\end{figure}

We show the results obtained for the R-NAMD and A-NAMD in Figs.~\ref{fig:namdr-core-sample} and \ref{fig:namda-core-sample}. Fig.~\ref{fig:namdr-core-sample} replicates the plot of R-NAMD vs. multiplicity as in T20 (their Fig.\,3) but with our expanded core sample of 67 systems vs. 13 in T20. 

Previous work~\citep{lim15} found an anti-correlation between the eccentricity and multiplicity of planetary systems. 
A subsequent study by \citet{zin17} linked this trend to the AMD-multiplicity trend. This result was made possible by the similar architectures of the considered systems, all of which were transiting. Thus, it was possible to directly compare across systems using the AMD metric. Finally, T20 re-observed this trend using the R-NAMD also in a larger sample of 99 systems where missing inclinations were constrained based on the equipartition of dynamical excitation from~\citet{las17}.

Fig.~\ref{fig:namdr-core-sample} shows the same anti-correlation between the R-NAMD and multiplicity with an expanded sample.

As already pointed out, the main limitation of the R-NAMD is that it does not capture the true architecture of the system, as it does not consider the spin-orbit angle. An example of how this does not capture the dynamical state of the system is provided by K2-290 and Kepler-462 we analyze in Sec~\ref{subsec:2sys}.
The R-NAMD, computed using relative inclinations between the planets, is insensitive to misalignment and therefore provides a lower limit to the dynamical excitation.
However, as it only requires photometric data, the R-NAMD can be computed for a larger sample of systems, see Figs.~\ref{fig:namdr-core-sample} and~\ref{fig:namda-core-sample} (spin-orbit angle data are still sparse). 
The knowledge of the spin-orbit angle, $\psi$, allows us to compute the A-NAMD, that measures the dynamical excitation corresponding to the true architecture of the planetary system.

Fig.~\ref{fig:namda-core-sample} shows the A-NAMD vs.~multiplicity.
Here, we used a sample of 18 systems\footnote{This was created from a custom database as not all the values were available on the NASA exoplanet archive. We provide the database at \texttt{ExoNAMD} GitHub under the data directory, saved as \texttt{custom\_db\_20250814.csv}.} where $\psi$ is known for the most massive planet or a planet of comparable mass. The multiplicity distribution in this sample is:

\begin{enumerate}
    \item 10 systems with $M$=2;
    \item 3 systems with $M$=3;
    \item 2 systems with $M$=4;
    \item 1 systems with $M$=5;
    \item 1 systems with $M$=6;
    \item 1 system with $M$=8 (the Solar System).
\end{enumerate}

When considering only the exosystems with known $\psi$ for all the planets (excluding the Solar System), we get a sample of only seven systems with $M$=2. 
We plot the 18 systems in Fig.\,\ref{fig:namda-core-sample} where we do not attempt to identify any trends in A-NAMD vs.~multiplicity given the scarcity of data points. However, we expect the eccentricity-multiplicity anti-correlation also to be present here. 

Finally, a trade-off between the number of characterized systems and the accurate physical description of their dynamical state can be reached by using projected spin-orbit angle, $\lambda$, instead of $\psi$ in Eq.~\ref{eq:namda}, which we call $\lambda$-NAMD. The advantage is that $\lambda$-NAMD does not require the knowledge of the stellar inclination.
The stellar inclination can be challenging to infer due to various astrophysical systematics, such as slow stellar rotation for M-dwarfs or a combination of several variability processes that can manifest for example in F-type stars (pulsations, ellipsoidal variations) that are over-imposed on the stellar rotation signal \citep{hen23}. 
The $\lambda$-NAMD can under- or over-estimate the dynamical excitation, but it provides a better approximation of the true dynamical state compared to that inferred from the R-NAMD.

\subsection{The case of K2-290 and Kepler-462}
\label{subsec:2sys}

\begin{table*}
\centering
\caption{Parameters of the K2-290 and Kepler-462 systems.}
\label{tab:k2290_k462}
\begin{tabular}{lcc|cc}
\toprule
 & \multicolumn{2}{c|}{K2-290} & \multicolumn{2}{c}{Kepler-462} \\
Parameter & Planet b & Planet c & Planet b & Planet c \\
\midrule
Inclination, $i$ [$^\circ$] 
  & $88.14^{+0.62}_{-0.50}$ & $89.37^{+0.07}_{-0.08}$ 
  & $89.34 \pm 0.05$ & $90.64 \pm 0.06$ \\
Eccentricity, $e$ & 0 & 0 & $0.056^{+0.019}_{-0.019}$ & $0.5^{+0.09}_{-0.09}$ \\
Semi-major axis, $a$ [au] & $0.5^{+0.09}_{-0.0}$ & $0.305^{+0.017}_{-0.017}$ & $0.473^{+0.021}_{-0.021}$ & $0.859^{+0.037}_{-0.037}$ \\
Projected spin-orbit angle, $\lambda$ [$^\circ$] 
  & $173^{+45}_{-53}$ & $153 \pm 8$ 
  & $-32 \pm 11$ & $-32 \pm 40$ \\
Spin-orbit angle, $\psi$ [$^\circ$] &  $117.8^{+14}_{-11}$  &  $124^{+6}_{-6}$  & $72^{+3}_{-3}$ & $73^{+11}_{-5}$ \\

\bottomrule
\multicolumn{5}{c}{System-wide NAMD values} \\
\midrule
$\log(\text{R-NAMD})$ 
  & \multicolumn{2}{c|}{$-5.17^{+0.30}_{-0.40}$} 
  & \multicolumn{2}{c}{$-1.07^{+0.11}_{-0.14}$} \\
$\log$($\lambda$-NAMD) 
  & \multicolumn{2}{c|}{$-0.03^{+0.01}_{-0.02}$} 
  & \multicolumn{2}{c}{$-0.84^{+0.29}_{-0.25}$} \\
$\log(\text{A-NAMD})$ 
  & \multicolumn{2}{c|}{$-0.11^{+0.02}_{-0.02}$} 
  & \multicolumn{2}{c}{$-0.44^{+0.04}_{-0.04}$} \\
\bottomrule
\end{tabular}
\end{table*}

\begin{figure*}[!h]
    \centering
    \includegraphics[width=0.32\linewidth]{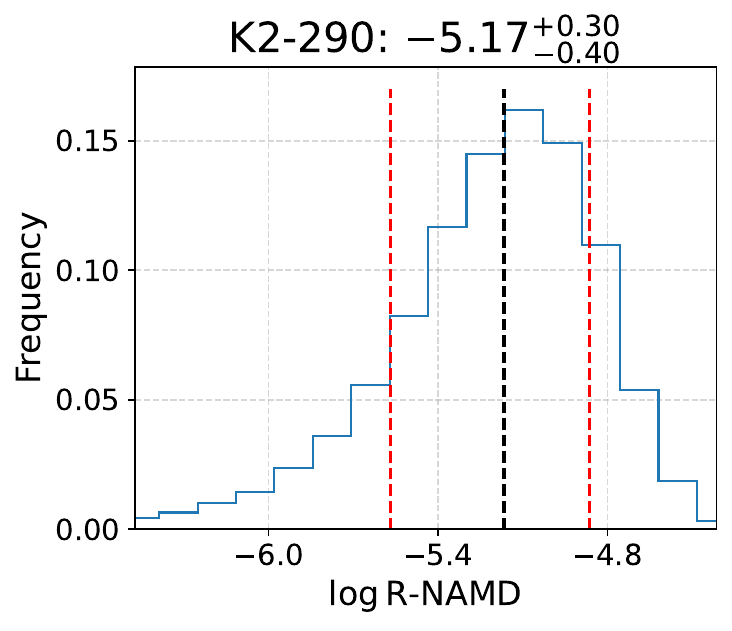}
    \includegraphics[width=0.32\linewidth]{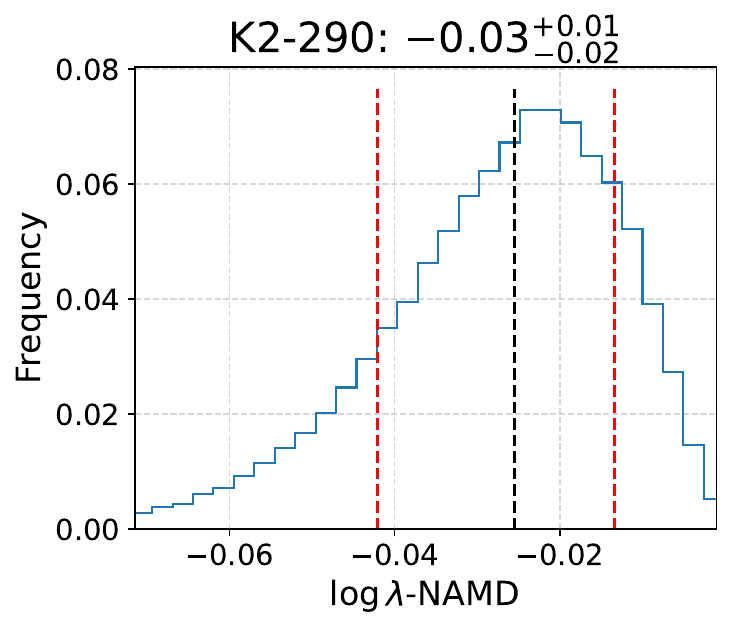}
    \includegraphics[width=0.32\linewidth]{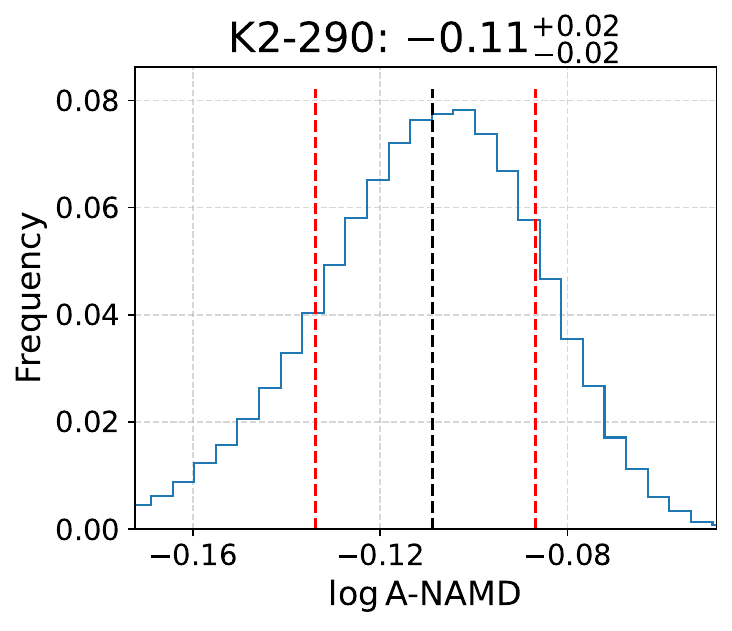}
    \caption{Comparison of NAMD values for the K2-290 system. The system consists of two planets on highly misaligned orbits. Left: R-NAMD as defined in T20 using relative inclinations. Middle: $\lambda$-NAMD, using projected spin-orbit angle ($\lambda$). Right: A-NAMD as defined in this work using the spin-orbit angle ($\psi$). The A-NAMD is the most suitable to capture the true architecture of the system as it is not affected by projection effects. }
    \label{fig:K2-290_namd}
\end{figure*}

In the following, we compare the various NAMD metrics on two systems, K2-290 and Kepler-462, whose parameters and computed NAMD values are listed in Table~\ref{tab:k2290_k462}. 
The K2-290 system hosts two planets on 9- and 48-day orbits \citep{hjo19}.
In Fig.~\ref{fig:K2-290_namd}, we compare the R-NAMD, A-NAMD, and $\lambda$-NAMD of this system, focusing on their respective advantages and on biases that are associated with their usage.

The R-NAMD value would be indicative of a quiescent history of the system. This contrasts with the observed misalignment of the planetary orbits that is instead reflected in the high value of A-NAMD. This high value indicates a high degree of dynamical violence during the history of the system and hints at a chaotic past, possibly driven by the two wide-companion stars \citep{best22}. Finally, the $\lambda$-NAMD overestimates the degree of dynamical violence by about 20\% compared to the A-NAMD, as the star is not seen fully edge-on and therefore $\psi$ is lower than $\lambda$ in this case (the qualitative information, however, is preserved).

Another instructive example is the Kepler-462 system which hosts two transiting exoplanets on well-separated but misaligned orbits \citep{ahl15}. The high R-NAMD value is mostly driven by the high eccentricities. The observed misalignment further increases the dynamical violence of the system and is reflected in the still higher A-NAMD. In this system, the $\lambda$-NAMD is 2.5 times lower than the A-NAMD (although it it consistent with the A-NAMD within 3--$\sigma$).
Similarly to K2-290, the difference between R-NAMD and A-NAMD is critical and highlights the usefulness of the A-NAMD to capture the true architecture of the system in an unbiased manner (see Fig. \ref{fig:Kepler-462_namd}).

\subsection{A new diagnostic: the four-quadrants diagram}
\label{subsec:4quad}

The A-NAMD metric introduced in this paper connects the information on both eccentricity and spin-orbit angle (or ``absolute inclination''). Both parameters provide a record of the dynamical state of the system (T22) and are closely linked together by the energy equipartition theorem~\citep{las17, he20}. Several mechanisms are responsible for exciting the planetary eccentricities and/or the spin-orbit angles and their effect over time manifests in the observed parameter space. So, by studying the distribution of the relevant dynamical parameters, and comparing them to theoretical model predictions, we can attempt to identify: (i) possible trends that emerge from these complex dynamical interactions, (ii) the most common evolutionary pathways, and (iii) the relative contributions of each pathway to the observed population. Ideally, these studies should be complemented by independent information from atmospheric studies that can reveal additional details and identify anomalies and knowledge gaps.

\begin{figure}[!h]
    \centering
    \includegraphics[width=1\linewidth]{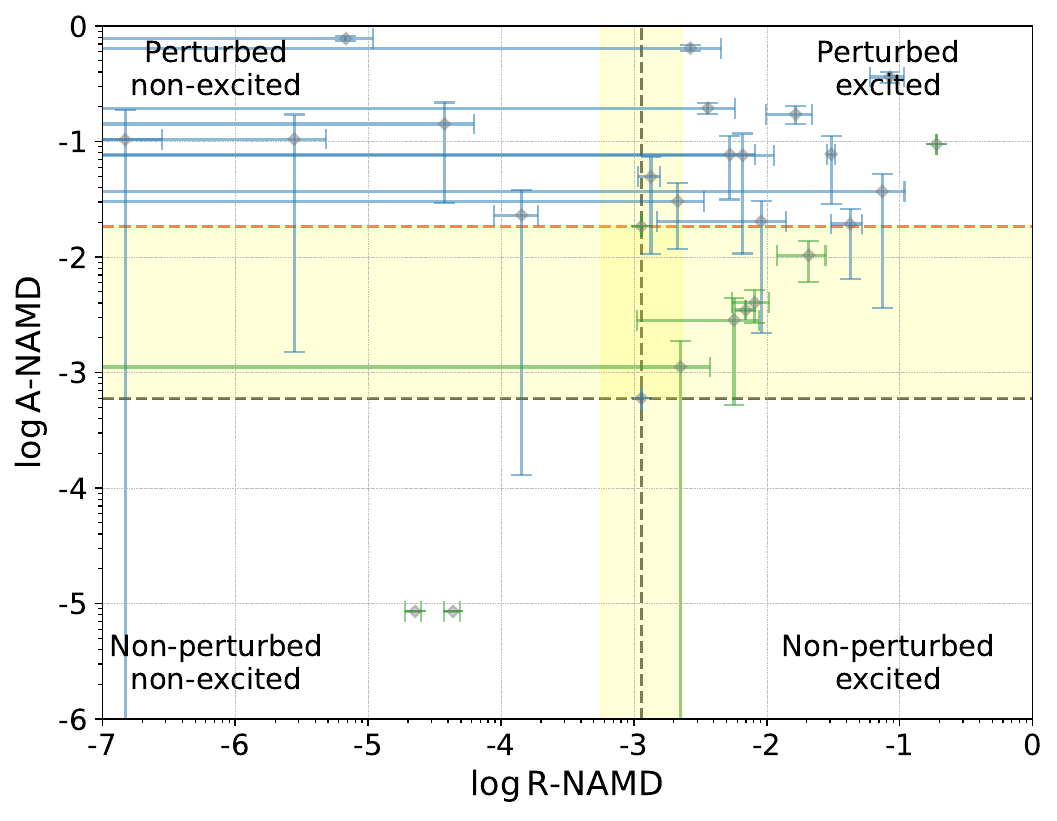}
    \caption{Four-quadrants diagram. This figure illustrates the possible states in which planetary systems can be found depending on their dynamical history which is reflected in their eccentricities and spin-orbit angles. The Solar System is located at the intersection of the two gray dashed lines and separates the four quadrants, as discussed in the text. As an illustration, we include a value (red line) of a Solar System with planets misaligned by 15 degrees. The vertical highlighted area marks instead the R-NAMD values that can be associated to excited but stable systems that did not undergo large-scale instabilities in the simulations by \citet{rick23}. In addition, we have included several synthetic systems based on real ones for which some parameters were missing (green points). These were inputted in order to better populate the quadrants. }
    \label{fig:fourquad}
\end{figure}

An example of a dynamical process where eccentricities and spin-orbit angles are tightly linked is high-eccentricity migration coupled with the Kozai-Lidov (KL) mechanism \citep{fab07}. In the KL mechanism, planets undergo periodic exchanges of eccentricity and spin-orbit angle with a third distant massive object and move inward towards the host star thanks to tidal forces. 
Another example which provides testable predictions is the formation of ultra-short-period (USP) planets.
\citet{mil20} suggested that USP planets should be aligned with the stellar spin axis, however, they should be misaligned with respect to the exterior planets.
Recently, \citet{yang25} suggested a new mechanism to explain the production of eccentric and highly misaligned hot Jupiters: the so-called ``eccentricity cascade''. This mechanism involves an outer companion star that induces eccentricity oscillations in the inner companion star through the KL mechanism, and these excitations are transferred to the planet through close encounters, triggering its migration toward the star. 
Between close encounters, the orbits evolve through secular diffusion~\citep{las17, he20}.

When discussing the interplay of eccentricities and spin-orbit angles, we will refer to systems characterized by eccentric orbits as ``excited'' and systems characterized by misaligned orbits as ``perturbed'' for reasons that will be clarified in the following. To quantify the prevalence of excitation and perturbation, we introduce a new diagnostic tool in the form of a diagram of A-NAMD vs.~R-NAMD. Figure~\ref{fig:fourquad} reports an example of this diagram, built with the 18 systems including the Solar System used in Fig.~\ref{fig:namda-core-sample}, and 9 synthetic systems\footnote{Located on the \texttt{ExoNAMD} GitHub under the data directory and saved as \texttt{custom\_db\_20250811\_fakes.csv}.} to populate the figure. 

Following T22, we used the Solar System values a dividing point in the diagram. The Solar System lies on the boundary between systems with orderly and chaotic evolution \citep{nes18}.
We also include a Solar System perturbed by 15 degrees as an illustration to account for a possible primordial misalignment during planet formation \citep{tak20}. 
Additionally, we include a yellow horizontal region between the Solar System and its version perturbed by 15 degrees. In this uncertainty region, planetary systems could be either perturbed or excited. Moreover, based on results from \citet{rick23}, we introduce another uncertainty region that extends vertically from half the R-NAMD of the Solar System to twice its value. Here, systems may or may not be excited.

The usage of the Solar System allows us to visually separate the following four quadrants:

\begin{itemize}
    \item \textbf{Top left}. Perturbed but non-excited systems, characterized by high A-NAMD and low R-NAMD values. These systems have misaligned orbits with low eccentricity, which could be indicative of past gravitational interactions that were more effective in the vertical direction, or stellar encounters that affected the disc before planets formed. In this quadrant, systems with short orbits might have had their orbits circularized but not yet realigned.
    \item \textbf{Top right}. Perturbed and excited systems, described by high A-NAMD and R-NAMD values. This is indicative of eccentric and misaligned orbits hinting at past gravitational encounters and highly violent dynamical history, possibly coupled with the KL mechanism or eccentricity cascade. Such systems are more likely to contain a distant companion on a wide orbit that might be responsible for the chaotic history, or have undergone a stellar encounter that destabilized the system.
    \item \textbf{Bottom left}. Non-perturbed and non-excited systems with low A-NAMD and R-NAMD values. These systems have likely experienced quiescent evolution, likely through disc migration, and were not altered by instability processes. For compact systems on short orbits, it is also plausible that their entire dynamical history has been erased through tidal forces.
    \item \textbf{Bottom right}. Non-perturbed but excited systems have low A-NAMD and high R-NAMD. This is indicative of eccentric and well-aligned orbits. This configuration can be reached by gravitational interactions that are more effective in the planar direction. For example, a distant companion that keeps a low but non-zero eccentricity of an inner companion without misaligning the orbits. Or, planet-planet scattering in a highly coplanar system.
\end{itemize}

These quadrants are further discussed in the next section, with a focus on current observational biases.

\section{Discussion}
\label{sec:discussion}

We have introduced the four-quadrants diagram as a new diagnostic tool to distinguish the prevalence of excitation and perturbation in the dynamical states of planetary systems. However, the interpretation of the four quadrants in Figure~\ref{fig:fourquad} needs additional context on possible observational biases that affect the currently observed population of exoplanets. Here, we summarize the main ones:

\begin{enumerate}
    \item The upper quadrants are currently over-populated. This is due to the errorbars on $\psi$ being on average 20 times larger compared to the median value of relative planetary inclinations for transiting exoplanets. These large errorbars drive up the A-NAMD values when using the Monte Carlo method explained before. However, with advances in instrumentation \citep{sei18,pepe21}, the $\psi$ parameter can be constrained to less than one degree \citep[e.g.,][]{casa17,wata24,rube24} with ultra-precise spectrographs, effectively correcting this bias. 
    \item Upper left quadrant. Old, compact systems will be over-represented in this quadrant as tidal forces act on faster timescales for orbit circularization compared to orbit realignment \citep{bono17}. This results in a faster right-left movement than top-bottom.
    \item Lower left quadrant. Only transiting planets (low relative inclinations) are suitable for spin-orbit angle measurements via the R-M effect, which provided the bulk of spin-orbit angle measurements so far. Thus, there is an observational bias to over-represent systems in this diagram compared to having a statistical sample of the general exoplanet population. However, this quadrant contains very few systems. A reason might be that the bias in point 1 is depopulating this quadrant and this issue will only be mitigated with more precise measurements of the spin-orbit angle.
    \item Another bias affecting the NAMD is the possibility of primordial misalignment. Recent works \citep[e.g.,][]{kuff24} have suggested that primordial misalignment can produce misaligned orbits by more than 15--20 deg, as originally suggested by \citet{tak20}. This creates an uncertainty region for the interpretation of the A-NAMD.
\end{enumerate}

A similar uncertainty region to that mentioned in point 4 affects also the interpretation of R-NAMD, as it is unclear what is the minimum excitation created by instabilities~\citep{tur22,rick23}.

As mentioned in the second point, tidal dissipation can severely affect the derived NAMD values. Hence, computing the dynamical timescales \citep[circularization and realignment timescales;][]{gold66, adam06} allows us to constrain and better interpret the observed position and possible movements across the four-quadrants diagram; this, in turn, allows us to correct the occurrence rates of planetary instabilities and star-triggered effects.

Tidal forces act on various timescales\footnote{Usually from 10 Myr and upwards.} on the spin orbit angle and planetary eccentricities, and affect the innermost planets more significantly. 
Therefore, the effect of tidal forces is most relevant for systems with planets on short-period orbits around cool stars with convective envelopes, for which the tidal forces are more efficient. Notably, the distribution of the spin-orbit angle for hot Jupiters is shaped both by gravitational and tidal effects \citep{and21,spal22}.
An illustrative example is provided by the shortest-period hot Jupiter in our sample of systems discussed in Fig.~\ref{fig:namda-core-sample}: WASP-47\,b. The timescales for the circularization and the realignment of its orbit  would be between 0.1 and 1 Gyr (depending on the tidal quality factor) and 1\,000 Gyr, respectively. Compared to the age of the system of $\sim$ 6--7 Gyr \citep{alm16}, the planetary eccentricity has been likely affected since its formation by the tidal forces while the spin-orbit angle has not.
So, we recommend that a comparison between the dynamical timescales and the age of the star be done especially for planets on short-period orbits to assess whether the dynamical information has been diluted over time~\citep{harr24}.
This is crucial to check for possible biases in the interpretation of the NAMD values when searching for population-level trends.

\subsection{Planetary atmospheres}
\label{SS:PA}

A distinct and complementary probe of the dynamical state is the composition of planetary atmospheres, which can help reveal past evolution even when the NAMD information is biased.
Planetary atmospheres are unique windows into the interior composition of planets, which holds the record of how and where they formed \citep{ob11,tur21,schne21}. At the same time, they connect the planet to its host star and are shaped by the stellar environment. Therefore, multiplanetary systems under the same stellar environment offer us opportunities to test planetary evolution and migration theories, e.g., the link between migration and composition, or the formation of compact architectures \citep{pace22}.
Specifically, the derived elemental ratios such as C/O, C/N, N/O, and S/N from spectroscopic measurements of exoplanet atmospheres, can help us distinguish between various planetary migration scenarios \citep{tur21,cross23} as they are strongly dependent on the initial semi-major axis of the planet orbit. 

Several mechanisms can alter the signatures of the different migration pathways \citep{daw18}. For instance, tidal forces can realign the originally misaligned planet. Or, the originally aligned planet can later be misaligned by some additional mechanisms \citep[e.g., internal gravity waves,][]{rog13}. To an extent, studying the atmospheric composition can lift the veil on previous history. 
Another example of the atmosphere-dynamics connection was presented in~\citet{seth25}. They found that misaligned planets tend to be more inflated (through tidal heating) than aligned ones, suggesting that these planets may have experienced dynamically violent processes in their past.
In this context, the introduction of the A-NAMD and the four-quadrants NAMD diagram as a new diagnostic tool, as well as the open-source code \texttt{ExoNAMD}, will streamline the assessment of the connection between atmospheric composition and dynamical state, providing the community of observers with additional context information to select the most suitable targets.

This atmospheric connection to planetary dynamics is being investigated in the ongoing BOWIE-ALIGN survey \citep{kirk24} which has selected a small sample of single-planet systems to follow-up with JWST: four aligned and four misaligned planets. The main objective of this survey is the comparison of the measured atmospheric composition between dynamically active and quiescent systems, searching for possible trends to be confronted by theoretical predictions \citep{pen24}.
Looking ahead, it is necessary to extend this kind of investigation to a much larger sample of systems, as well as include multiplanetary systems to conduct comparative planetology studies under the same stellar environment. 
In this regard, the data from the Ariel space mission will be invaluable for studying population-level trends. 
Ariel\footnote{\href{https://sci.esa.int/web/ariel/-/ariel-definition-study-report-red-book}{Ariel Definition Study Report (Red Book)}} aims to perform the first unbiased spectroscopic survey of hundreds of exoplanetary atmospheres
in the visible and infrared~\citep{Tinetti2018}. Its continuous coverage from 0.5 to 7.8 micron, observed in one shot, and excellent stability at L2~\citep{boch25, boch25b}, makes it the go-to mission for large-scale homogeneous exoplanet studies.

NAMD may be included as a parameter in the Ariel candidate target list~\citep{Edwards2022} to aid in selecting an unbiased, statistical sample of planets, also from the dynamical point of view. The interpretation of the Ariel observations could then be contextualized using the dynamical information provided by the NAMD~\citep{tur22}, as computed by our \texttt{ExoNAMD} tool. In this way, the results of the pathfinder BOWIE-ALIGN survey will be expanded to a statistical sample of hundreds of exoplanets, characterized homogeneously by Ariel. 

The importance of having homogeneous stellar and planetary parameters has been recognized by the Ariel mission consortium as a top priority to obtain an unbiased picture of the parameter space of exoplanets. This led to the formation of various working groups. Today, all parameters needed to compute the NAMD have their respective WGs already in operations (e.g. Stellar Characterization WG\footnote{\url{https://sites.google.com/inaf.it/arielstellarcatalogue/}}), including the Stellar Obliquity WG\footnote{\url{https://arielmission-stellarobliquity.github.io/}} which aims to provide spin-orbit angle measurements for the majority of mission target candidates \citep{zak25b} with a special focus on Tier-3 candidate targets and multiplanetary systems. The \texttt{ExoNAMD} tool, developed within the Ariel Stellar Obliquity WG, is open-source and we envisage its usage by the community of observers and its adaptation to the needs of other observatories to aid in target selection and data interpretation.

\subsection{Future outlook}

Currently, the systems for which we possess all the needed dynamical parameters for the NAMD calculation cover only a limited portion of the parameter space of the exoplanetary population. This precludes us from having a comprehensive picture of possible shaping mechanisms and it also makes it challenging to isolate the common processes affecting all subpopulations.
Here we comment on the ongoing and upcoming missions that will provide new windows into under-explored populations.

A currently under-explored population is that of planets on wide orbits, such as cold Jupiters. A way to isolate the effects of migration on the population of Jupiters would be to assess the dynamical excitation of hot and warm Jupiters vs. cold Jupiters, as the latter migrated far less or remained close to the region of formation.
The number of systems with planets on wide orbits for which the NAMD can be determined is still limited, but has seen steady growth in recent years. Thanks to astrometry and direct imaging methods \citep[e.g.,][]{stol25}, we have been able to increase the detection of long-period planets as well as constrain their spin-orbit angles.
The upcoming DR4 Gaia release is expected to provide additional planets on wide ($>$ 1\,au) orbits. While for companions at wide separations it is difficult to measure the spin-orbit angle, lower limits can be inferred by comparing the stellar rotation velocities and orbital inclinations. \citet{bowl23} found that companions between 10 and 250\,au are consistent with having a spin-orbit angle distribution with a mixture of isotropic and aligned systems. 

The \textit{PLATO} mission \citep{rauer24} is expected to discover between 4\,600 and 24\,000 exoplanets during its nominal mission depending on the assumed planetary model \citep{mat23}. Compared to the \textit{TESS} mission \citep{rick14} it is designed to yield planets on longer orbital periods (50--200\,days), significantly expanding our current knowledge of the exoplanetary population.
The about 700 multiplanetary systems known from the \textit{Kepler} mission \citep{koch10} mostly orbit faint stars (median magnitude is 14.5), making follow-ups rather difficult. \textit{PLATO} is expected to yield at least the same number of multiplanetary systems around brighter stars than those discovered by \textit{Kepler}.
Several ground-based high-resolution spectrographs~\citep[e.g., PLATOSpec;][]{kab25} will dedicate a large fraction of their available time to follow-up \textit{PLATO} targets with the aim of measuring their masses, eccentricities and spin-orbit angles. Hence, the rapidly growing number of systems with measured spin-orbit angles is expected to grow even faster as well as their precision. The simultaneous photometric and spectroscopic observations can provide more precise spin-orbit angle measurements than individual spectroscopic measurements as there are joint parameters in the light curve and the R-M effect analysis.
Furthermore, multiple systems discovered by \textit{PLATO} will be characterized by transit-timing-variations (TTVs) with higher precision than by \textit{Kepler}, thanks to the longer observation time and higher sampling rate, providing meaningful constraints to the processes of planetary formation and evolution, especially in regions of the parameter space beyond the reach of previous missions\footnote{\href{https://sci.esa.int/s/8rPyPew}{\textit{PLATO} Definition Study Report (Red Book)}.}.

Another mission relying on the TTV characterization of exoplanets is the Nancy Grace Roman Space Telescope~\citep{ake19}. Roman is the next NASA flagship mission that aims to detect and characterize exoplanets in the near-infrared. The predicted exoplanet yield ranges from 60 to 200 thousand transiting planetary candidates, and from 7 to 12 thousand small transiting exoplanets \citep{wilson23}. The main detection method will be through transit detection, but Roman is expected to find several multiplanetary systems also with microlensing, further expanding the planetary parameter space to larger orbital distances \citep{penny19,fat23}.
Most of the multiplanetary systems discovered by Roman will be fainter compared to \textit{Kepler} targets and hence they will not be ideal for ground-based follow-up. Thus, the spin-orbit angle parameter will remain largely unconstrained. In the absence of spin-orbit angle, the R-NAMD can still be used for dynamical characterization, with the caveats described before.

The next-generation high-resolution ultra-stable instruments -- including ANDES \citep{pall23} at ELT, iLocater \citep{crep16} at LBT, and G-CLEF \citep{szen16} at GMT and many others -- will significantly increase the number of spin-orbit angle measurements, covering the parameter space even of smaller planets. This will unlock the A-NAMD calculation and the dynamical assessment within the four-quadrants diagram even for these planets which were previously inaccessible, e.g., due to the small R-M effect amplitude or long transit duration requiring multiple of these instruments. 
 
\section{Summary}

The exoplanetary research community faces three critical challenges: first, selecting optimal targets for detailed characterization with telescopes like JWST; second, compiling unbiased target lists for missions such as Ariel that aim to statistically explore the parameter space of the planetary population. The third challenge is correctly interpreting the atmospheric data \citep[Three Key Criteria,][]{seag25}, for which the context information on the planetary system and its evolution, including dynamical clues, is critical (T22).

To aid in addressing these challenges, this work introduces the A-NAMD metric as a powerful probe of the past dynamical histories of multiplanetary systems. This metric incorporates the spin-orbit angle parameter, which has been measured for a growing number of planets in recent years, to capture the true 3-D architecture of the system. For individual system studies, the A-NAMD can be used in synergy with existing stability metrics such as SPOCK, MEGNO, and AMD, which provide information on the future dynamics whereas NAMD provides information on the past behavior.

We demonstrated the diagnostic power of the joint use of the A-NAMD and the previous NAMD formulation (R-NAMD) by combining them into a 4-quadrants diagram, enabling qualitative classification of systems' dynamical histories. This diagram enables distinguishing between the systems where dynamical events primarily affected the planar direction of the planetary orbit (producing eccentric systems) versus those where vertical effects dominated (resulting in misaligned systems). We set the Solar System, with its well-constrained parameters, as a reference point for NAMD values.

The planetary community heavily relies on the stability metrics and their calculation is implemented, for instance, in the open-source \texttt{REBOUND} software~\citep{rein12}. These metrics concern the future stability of multiplanetary systems and our work seeks to complement them with characterization of the past dynamical history. Despite the fact that the R-NAMD was widely adopted by the community, there is no open-source tool to calculate it.
Thus, to directly support the community, we developed \texttt{ExoNAMD}, a user-friendly Python tool that calculates both the R-NAMD and our newly introduced A-NAMD. This open-source package features comprehensive documentation and can be installed directly from PyPI\footnote{\url{https://pypi.org/project/exonamd/}} or GitHub\footnote{\url{https://github.com/abocchieri/ExoNAMD/}}. Its flexible API allows users to easily customize calculations with their own data, for example, by adapting the example input databases given in the documentation\footnote{See under the exonamd/data directory in the GitHub repository.}.

Our envisioned usage of the A-NAMD metric is to provide context information for building target lists for upcoming atmospheric missions, and facilitate optimal target selection by observers\footnote{The NAMD value provides dynamical context rather than target observability such as Transmission Spectroscopy Metric \citep[TSM,][]{kempton2018}}. In this regard, we remind that NAMD metrics represent a lower boundary on the intensity of past dynamical events, as processes like tidal interactions erase evidence of past events over time. To reconstruct the complete evolutionary pathways of planetary systems, we must combine dynamical clues -- for instance those provided via NAMD analysis -- with detailed atmospheric composition studies. Atmospheres can preserve crucial evidence of formation and evolution processes, enabling us to develop a more complete and robust understanding of planetary systems' histories.

\begin{acknowledgements}
A.B. and D.T. are supported by the Italian Space Agency (ASI) with \textit{Ariel} grant n. 2021.5.HH.0. 
J.Z. acknowledges the support from GACR:22-30516K. J.Z. and D.T. acknowledge support from the COST Action CA22133 PLANETS. D.T. acknowledges support from the European Research Council via the Horizon~2020 framework program ERC Synergy ``ECOGAL'' project GA-855130.
This research was supported by the Munich Institute for Astro-, Particle and BioPhysics (MIAPbP), which is funded by the Deutsche Forschungsgemeinschaft (DFG, German Research Foundation) under Germany´s Excellence Strategy – EXC-2094 – 390783311. We are grateful to Henri Boffin, Dominika Itrich and Pavol Gajdos for the insightful discussion improving this work.
\end{acknowledgements}

\bibliographystyle{aa}
\bibliography{aa}

\begin{appendix}

\section{NAMD parameters}
\label{app:params}

Here we list the NAMD parameters and the most common ways to obtain them. For a comprehensive review see \citet{perr18}.

The planetary mass, $m_k$, is most commonly obtained with the radial velocity method using high-resolution spectrographs \citep{wr18}. Another method of obtaining planetary masses is the TTV method \citep{agol18}, which is mainly relevant for multiplanetary systems in or near resonant configurations.

The semi-major axis, $a_k$, can be derived by fitting a transit light curve or a radial velocity curve\footnote{For consistency with previous works by CH1 and T20 we use $a_k$ instead of $P_k$.} provided that the stellar mass is known.

The eccentricity, $e_k$, can be derived both from photometry \citep{daw12} or spectroscopy using the radial velocities. N-body simulations can be used to derive upper limits for eccentricities in multiplanetary systems under the assumption that the systems are stable. Even very stable systems such as the resonant chain of six planets HD 110067 can have non-zero eccentricity in their current architecture. For example, \citet{lam24} have shown that this system can be stable with planetary eccentricities around 0.1.

The planetary inclination, $i$, is usually obtained from fitting the transit in the light curves and subsequently calculated from the derived impact parameter. 

The stellar inclination, $i_*$, can be derived by measuring the rotational modulation \citep[e.g., by stellar spots,][]{ska22}.
\citet{mas20} have shown that a simple calculation of $i_*$ can lead to biases as the rotational velocity $v$ and \(v sin\,i_*\) are not independent quantities; hence, a Bayesian approach is recommended to derive $i_*$.

An alternative approach to derive $i_*$ employs asteroseismology. Due to stellar rotation, splitting of pulsation modes \citep{aer21} allows us to independently measure the $i_*$ parameter. \textit{Kepler} data has been used to obtain the majority of such measurements \citep{hub13, camp16}; many more are envisioned to be obtained from the upcoming \textit{PLATO} mission. Together with dedicated ground-based spectrographs measuring the projected spin-orbit angle, $\lambda$, this will provide a rich dataset of spin-orbit angle values. 

Finally, the spin-orbit angle, $\psi$, is obtained by combining $\lambda$ derived from the R-M effect, the planetary inclination, $i_p$, and the stellar inclination, $i_*$, using the spherical law of cosines:

\begin{equation}
\cos\psi = \sin i_*\, \sin i \, \cos|\lambda| + \cos i_*\, \cos i.
\end{equation}

\section{Missing parameters}
\label{app:missing}
When the complete set of parameters for the NAMD calculation is not available (e.g., not existing in the NASA catalog), the user can pass their own database to \texttt{ExoNAMD}. Examples are in the GitHub\footnote{\url{https://github.com/abocchieri/ExoNAMD/}} repository under the \texttt{data} folder. Alternatively, \texttt{ExoNAMD} provides several options to interpolate missing values. In this case, it is recommended to apply caution before interpreting the resulting NAMD, as it may rely on undesired/wrong assumptions. In the following paragraphs, we detail how each missing parameter is interpolated by default so that the user can decide if the assumptions are appropriate for their science case. 

\paragraph{$\psi$.} To date, only seven multiplanetary systems have the $\psi$ measured for all the planets; for the remainder, we can input the missing values if $\psi$ of the most massive planet is available. In this case, we assume the same value of $\psi$ for all the other planets. 
This approximation is justified by the conservation of angular momentum w.r.t. the most massive planet, as argued by T20 for the R-NAMD. However, caution must be exercised, as we now have evidence from both N-body simulations and $\psi$ measurements confirming the existence of multiplanetary systems with planets with vastly different $\psi$ even for compact system architectures \citep{bou21, yu25}. We note that this further motivates the measurements of $\psi$ for multiple planets in a given system.

\paragraph{$m_\mathrm{p}$.} Large photometric surveys have produced a myriad of planetary candidates without firm planetary mass measurements. The mass can be approximated using carefully calibrated $M$-$R$ relations \citep[e.g.,][]{parvi24,mull24} that were designed for various types of planets (e.g., Super-Earths, Sub-Neptunes, Jupiters). 
In \texttt{ExoNAMD}, if the mass of the planet is missing, and both the planetary radius and the associated uncertainties are available, the default setting is to obtain the planetary mass using the Bayesian $M$-$R$ relation implemented in the open-source \texttt{spright}\footnote{\url{https://github.com/hpparvi/spright}} tool \citep{parvi24}. The calibrated range of planetary radii is [0.5, 6] \(R_{\oplus}\). For any radius outside this range, the mass will not be interpolated. We have decided to adopt the \texttt{spright} tool given that most of the known multiplanetary systems have planets with radii between these values. 

\paragraph{$e$.} \citet{zin17} have shown there is an average eccentricity, $e$, vs. planetary multiplicity, $M$, anti-correlation and they derived a relation between the two parameters. If $e$ is not available, \texttt{ExoNAMD} will approximate it using this relation by default. In some cases, it is advisable to input an eccentricity value (or its upper limit) derived from an N-body simulation, from which an upper limit of the NAMD can be obtained.

\paragraph{$i$.} When the planetary inclinations are not available for some planets \texttt{ExoNAMD} assumes that the planets are on co-planar orbits w.r.t. the most massive planet in the system. If this planet does not have its inclination measured, the default is to assume the inclination of the next most massive planet that has its inclination measured for the rest of the planets. When no inclination is available for any planet in the system, we assume $i$ = 90$^{\circ}$ and coplanarity to increase the sample.

\paragraph{$i_*$ or $\lambda$.}
In case of missing stellar inclination, $i_*$, or projected spin-orbit angle, $\lambda$, the $\psi$ values cannot be determined; therefore, only the R-NAMD can be computed. The user may wish to use $\lambda$ when $\psi$ is not available; in Sec.~\ref{sec:results}, we discuss the possible biases that this choice may entail.

\paragraph{Missing uncertainties.}
The general rule for interpolating missing uncertainties is to set them to 0. Although this may result in some biases, we adopted this methodology to increase the sample. We discuss the resulting bias in the following paragraph.

\paragraph{Flags}

To keep track of the interpolated values, we introduce a system of descriptive flags that allows full traceability of each interpolated value for the parameters used in the NAMD calculation. 
Table~\ref{tab:flag_parameters} summarizes the flags corresponding to each interpolated planetary parameter.

\begin{table}[!h]
    \centering
    \caption{Description of flags used in \texttt{ExoNAMD}.}
    \begin{tabular}{cl}
        \toprule
        \textbf{Flag} & \textbf{Parameter} \\
        \midrule
        1 & Eccentricity \\
        2 & Mass \\
        3 & Inclination \\
        4 & Semi-major axis \\
        5 & Spin-orbit angle \\
        - & Associated lower errorbar \\
        + & Associated upper errorbar \\
        d & Do not use \\
        \bottomrule
    \end{tabular}
    \label{tab:flag_parameters}
\end{table}

Each planet starts with "0" as the flag and then for each interpolation, the above flags are appended as necessary, in an interpretable way. For instance, if we interpolated the eccentricity and the spin-orbit angle, together with their uncertainties, the resulting flag would be [01+-5+-]. Flags are stored automatically in the database produced by \texttt{ExoNAMD}. If ``d'' is present in the flag, the parameter is absent for all planets in the system and we mark it as a ``do not use'' system.

As stated before, missing errorbars are interpolated by setting them to zero by default to keep more targets in the sample. As a consequence, the resulting NAMD values from our Monte Carlo procedure provide a lower limit by definition. This artifact is most prominent when the errorbars of the eccentricity and spin-orbit angle are missing. Hence, we recommend to check the flags to interpret any results obtained with interpolated values. 

\section{Additional Figures}

\begin{figure*}[!h]
    \centering
    \includegraphics[width=0.32\linewidth]{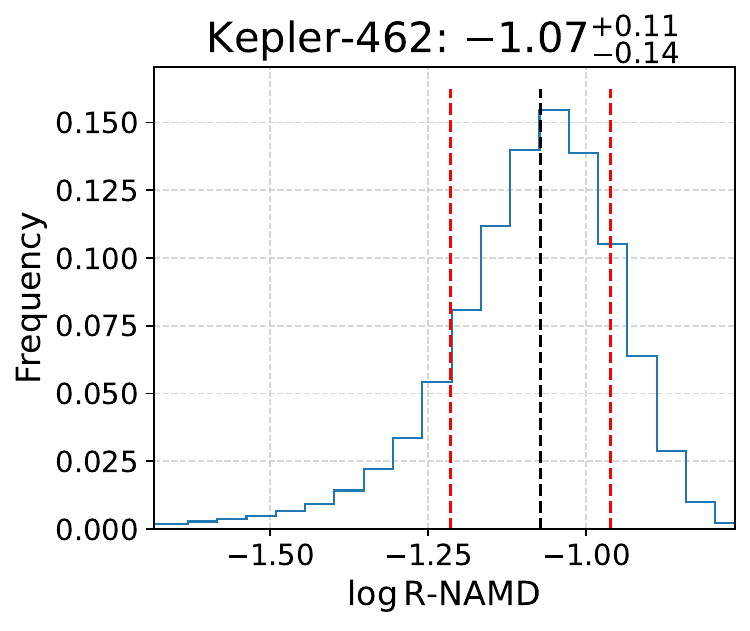}
    \includegraphics[width=0.32\linewidth]{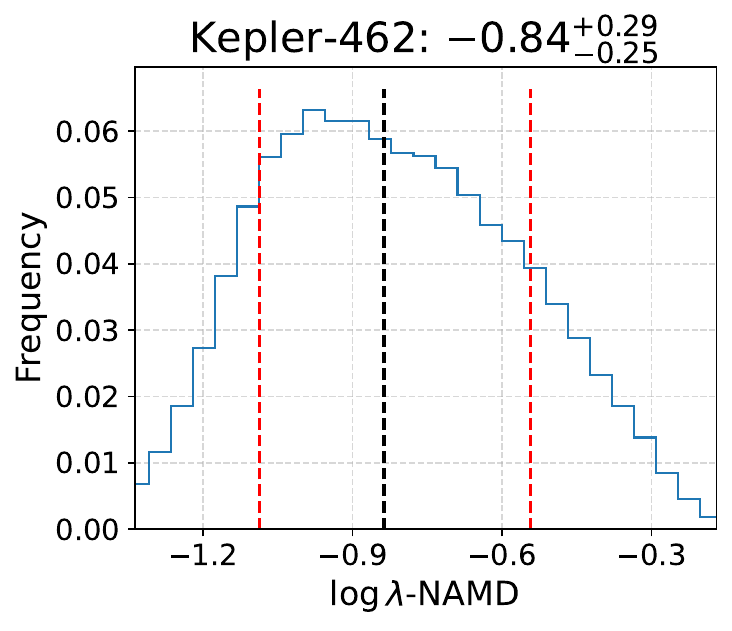}
    \includegraphics[width=0.32\linewidth]{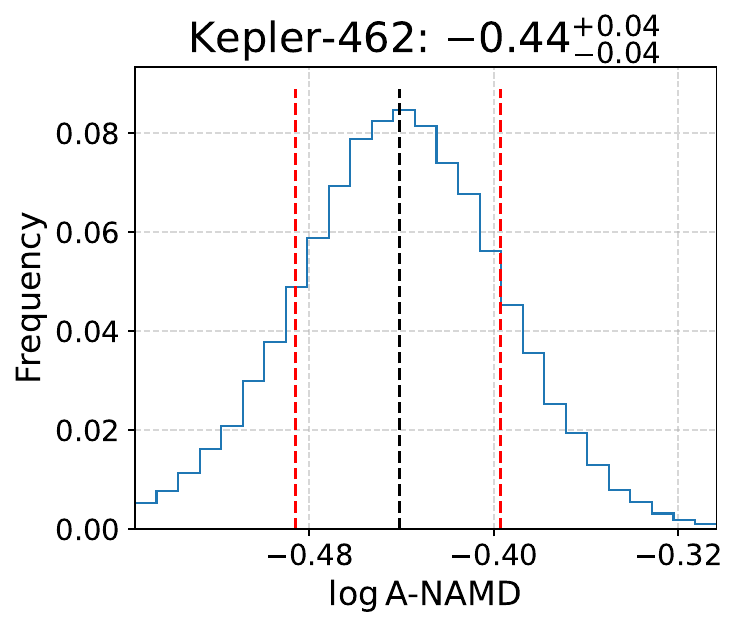}
    \caption{Comparison of NAMD values for the Kepler-462 system. The system consists of two planets on eccentric and misaligned orbits. Left: R-NAMD as defined in T20 using relative inclinations. Middle: $\lambda$-NAMD, using projected spin-orbit angle ($\lambda$). Right: A-NAMD as defined in this work using the spin-orbit angle ($\psi$). The A-NAMD metric is the most suitable to capture the true architecture of the system. }
    \label{fig:Kepler-462_namd}
\end{figure*}

\end{appendix}
\end{document}